\documentclass[doublecol]{epl2}

\usepackage{amsmath,amssymb}

\title{Observational Viability of the \textbf{Intermediate} DBI Inflation in the Presence of a Minimal Length}

\author{N. Rashidi\inst{1,2}$^{}$\footnote{n.rashidi@umz.ac.ir (Corresponding Author)} \and
M. Roushan\inst{1,2}$^{}$\footnote{m.roushan@umz.ac.ir}
 \and K. Nozari\inst{1,2}$^{}$\footnote{knozari@umz.ac.ir}}
\shortauthor{N. Rashidi \etal}

\institute{ \inst{1} Department of Theoretical Physics, Faculty of
Science,
University of Mazandaran,\\
P. O. Box 47416-95447, Babolsar, IRAN\\
  \inst{2} ICRANet-Mazandaran, University of Mazandaran,\\
P. O. Box 47416-95447, Babolsar, IRAN }

\abstract{ We consider an intermediate Dirac-Born-Infeld (DBI)
inflationary model in the presence of a minimal measurable length in
the theory. We show that, the presence of a minimal measurable
length modifies the definitions of the scalar and tensor spectral
indices and also other inflation observables. This is due to
modification of the momentum and corresponding wave number of the
perturbations in the presence of a minimal length. By using the
deformed definition of the scalar and tensor spectral indices, we
perform numerical analysis on the intermediate DBI inflation model
to find some constraints on the deformation parameter. In this
regard, we compare our numerical results with both Planck2018 TT,
TE, EE +lowE +lensing +BAO+ BK14 and Planck2018 TT, TE,EE
+lowE+lensing+BK14 +BAO+LIGO $\&$ Virgo2016 data at the $68\%$ CL
and $95\%$ CL. Our numerical study shows that the intermediate DBI
inflation model in the presence of a minimal measurable length is
observationally viable if the upper bound on the deformation
parameter to be considered of the order of $10^{48}$ at $68\%$ CL
and $10^{49}$ at $95\%$ CL. This is consistent with the results of
other approaches to constrain such a quantity.
  }

\begin{document}

\maketitle

\section{Introduction}

Given that the standard model of cosmology suffers from some
problems, cosmologists have tried to solve these problems by
introducing the inflation paradigm. The inflation, in its simplest
form, is caused by a so-called ``inflaton" field which is a
canonical single scalar field. By slow-rolling of this field on its
almost flat potential, the early universe experienced an exponential phase of
expansion. This inflationary expansion, which is supposed to have been
occurred in the very early universe, can address some of the problems of the
standard model. Although the simple single field model predicts
scale-invariant, adiabatic and Gaussian dominant mode of the
primordial perturbations, there are many inflation models which
predict the dominant modes of the perturbations to be non-Gaussian
distributed~\cite{Gut81,Lin82,Alb82,Lin90,Lid00a,Lid97,Rio02,Lyt09,Mal03}.
One of these models is the DBI (Dirac-Born-Infeld)
inflation~\cite{Sil04,Che07,Ali04,Che05,Noz13,Amani18, Rasouli19,Noz19,Ras21a}. A DBI
field in the string theory is defined as the radial position of a D3
brane. Cosmologist have shown that in the inflationary models based
on the DBI field, the non-Gaussianity of distribution of perturbations is
possible.

On the other hand, since in the early universe we deal with high
energy regime, incorporating the quantum effects seems logical and
in fact inevitable. In this regard, authors of
Refs.~\cite{Duf97,Noj02,Noj03,Ras21} have studied the inflationary
phase of the universe in the presence of quantum effects. However,
in this work, we include the quantum effects in a different way. In
fact, we consider the inflationary phase in the presence of a
natural cutoff. In this respect, natural cutoffs are constraints
that have appeared by the merging of the two respected theories;
quantum mechanics and general relativity, as a unified theory.
Therefore, considering natural cutoffs as a necessity in the very
nature of spacetime leads to fundamental modifications in the
structure of the standard quantum mechanics. It has been designated
that gravity in very small length scales makes a consequential
modification in the fabric of spacetime. These modifications induce
minimal uncertainty in the positions of the particles in atomic and
subatomic levels
~\cite{Venez86,Amati87,Gross87,Amati89,Konishi90,Mag93,Capo2000,
Magg93,Mag94,Hossen2012,Hossen2013,Scar1999}. Actually, there exists
minimal uncertainty in the measurement of the position of a test
particle in a quantum mechanical system. The essence of this
intrinsic cutoff demands deformity of the standard Heisenberg
Uncertainty Principle to the Generalized Uncertainty Principle (GUP)
(see, for more more details
~\cite{Kempf95,Nozari12b,Roushan14,Roushan16,Roushan18,Roushan19}).

In this paper we investigate the effects of the existence of a minimal length as an
inherent natural cutoff in the fabric of spacetime, on a specific
inflation model within a distinctive approach. Given that, in the
presence of the minimal length, the momentum is modified as
$p^i\rightarrow p^{i}(1+\alpha p^2)$, the corresponding wave number
also is modified as $k^i\rightarrow k^{i}(1+\alpha k^2)$. This
modification in wave number shows its effects in the definition of
the scalar and tensor spectral indices as well. In this way, it
seems that the presence of the minimal length can affect the
observational viability of the inflation models or at least changes the viable
ranges of the model's parameter space. To check the observational
viability of our model, we use Planck2018 data sets. For instance,
the Planck2018 TT, TE, EE+lowE+lensing+BAO+BK14 data, based on the
$\Lambda CDM+r+\frac{dn_{s}}{d\ln k}$ model, implies a constraint on
the value of the scalar spectral index as
$n_{s}=0.9658\pm0.0038$~\cite{pl18a,pl18b}. The constraint on the
tensor-to-scalar ratio, obtained from the same data set is
$r<0.072$~\cite{pl18a,pl18b}. There is also a constraint on the
tensor spectral index, obtained from Planck2018 TT, TE, EE
+lowE+lensing+BK14+BAO+LIGO and Virgo2016 data as
$-0.62<n_{T}<0.53$~\cite{pl18a,pl18b}. By comparing the numerical
results of our analysis in this model with the mentioned
observational data, we find some constraints on the model
parameter's space, specially the deformation parameter.

In this regard, this paper is organized as follows. In section 2, we
review the machinery of the DBI inflation briefly. In this section, we present the
background equations, slow-roll parameters, and some important
perturbation parameters. In section 3, we reconstruct the model and
obtain the slow-roll parameters in terms of the Hubble parameter.
In section 4, the effects of the presence of a minimal length are
studied. In section 5, we consider an intermediate DBI model and
study the model in the presence of a minimal measurable length. In
this section, we also perform numerical analysis on the model and
compare the result with both Planck2018 TT, TE, EE +lowE +lensing
+BAO+ BK14 data and Planck2018 TT, TE, EE +lowE+lensing+BK14
+BAO+LIGO $\&$ Virgo2016 data at $68\%$ CL and $95\%$ CL. In section
6, we present a summary of our work.

\section{DBI Inflation}

The action for the DBI model is as follows
\begin{equation}
\label{eq1} S= \int
d^{4}x\sqrt{-g}\Bigg[\frac{1}{2\kappa^{2}}R-f^{-1}(\phi)\,\sqrt{1-2f(\phi)X}-V(\phi)
\Bigg],
\end{equation}
where, $R$ is the Ricci scalar,
$X=-\frac{1}{2}\partial_{\nu}\phi\,\partial^{\nu}\phi$, and the
potential of the field is denoted by $V(\phi)$. Also,
$f^{-1}(\phi)$ is the inverse of the brane tension. This parameter in the
DBI model is related to the geometry of the throat of the compact
manifold which is passed by the D3-brane.

When we vary the action (\ref{eq1}) with respect to the metric, we
obtain Einstein's field equations as
\begin{eqnarray}
\label{eq2}
G_{\mu\nu}=\kappa^{2}\Bigg[-g_{\mu\nu}f^{-1}\sqrt{1+f\,g^{\mu\nu}\,\partial_{\mu}\phi\,\partial_{\nu}\phi}-g_{\mu\nu}V
\nonumber\\+\partial_{\mu}\phi\,\partial_{\nu}\phi
\Big(1+f\,g^{\mu\nu}\,\partial_{\mu}\phi\,\partial_{\nu}\phi\Big)^{-\frac{1}{2}}\Bigg]\,.
\end{eqnarray}
In a spatially flat FRW background, characterized by the following metric
\begin{equation}
\label{eq3} ds^{2}=-dt^{2}+a^{2}(t)\delta_{ij}dx^{i}dx^{j}\,,
\end{equation}
Einstein's field equations give the following Friedmann equations
\begin{eqnarray}
\label{eq4}
3H^{2}=\kappa^{2}\Bigg[\frac{f^{-1}}{\sqrt{1-f\dot{\phi}^{2}}}+V\Bigg]\,,
\end{eqnarray}

\begin{eqnarray}
\label{eq5}
2\dot{H}+3H^{2}=\kappa^{2}\Bigg[f^{-1}\,\sqrt{1-f\dot{\phi}^{2}}+V\Bigg]\,.
\end{eqnarray}
The equation of motion of the scalar field in the DBI model is given by
\begin{equation}
\label{eq6}\frac{\ddot{\phi}}{(1-f\dot{\phi}^{2})^{\frac{3}{2}}}+\frac{3H\dot{\phi}}{(1-f\dot{\phi}^{2})^{\frac{1}{2}}}
+V'=-\frac{f'}{f^{2}}
\Bigg[\frac{3f\dot{\phi}^{2}-2}{2(1-f\dot{\phi}^{2})^{\frac{1}{2}}}\Bigg]\,,
\end{equation}
which is obtained by variation of the action (\ref{eq1}) with
respect to $\phi$.

To study the inflationary phase derived by the slow-rolling of the
field on its potential, the following slow-roll parameters are
defined
\begin{equation}
\label{eq7}\epsilon\equiv-\frac{\dot{H}}{H^{2}}\,,\quad
\eta=\frac{1}{H}\frac{d \ln \epsilon}{dt}\,,\quad
s=\frac{1}{H}\frac{d \ln c_{s}}{dt}\,,
\end{equation}
where $c_{s}$ is the sound speed of the perturbations and is given
by
\begin{eqnarray}
\label{eq8} c_{s}^{2}=1-f\dot{\phi}^{2}\,.
\end{eqnarray}

The inflationary expansion is obtained by considering
$f\dot{\phi}^{2}\ll 1$ and $\ddot{\phi}\ll3H\dot{\phi}$
(corresponding to $|\epsilon,\,\eta,\,s| \ll 1$). Now, the slow-roll
parameters in the DBI model take the following forms
\begin{equation}
\label{eq9}\epsilon\simeq{\frac {{f}^{2}{{\it
V'}}^{2}}{2{\kappa}^{2} \left( Vf+1 \right) ^ {2}}}-{\frac {{\it
V'}\,{\it f'}}{{\kappa}^{2} \left( Vf+1 \right) ^{2 }}}+{\frac
{{{\it f'}}^{2}}{2{f}^{2}{\kappa}^{2} \left( Vf+1 \right) ^{2}}} \,,
\end{equation}
\begin{equation}
\label{eq10}\eta\simeq -{\kappa}^ {-2} \left(  \frac{\left( 2\,{\it
V''}-2\,{\frac {{\it f''}}{{f}^{2}}} \right)}{ \left( V+{f}^{-1}
\right) }-\frac{ \left( {\it V'}-{\frac {{\it f'}}{{f }^{2}}}
\right) ^{2}}{ \left( V+{f}^{-1} \right)} \right)  \,.
\end{equation}
and
\begin{equation}
\label{eq11}s\simeq-\frac{f'\left(\frac{f'}{f^{2}}-V'\right)^{2}}{18\left(\frac{\kappa^{2}}{3}(f^{-1}+V)\right)^{\frac{3}{2}}}
\end{equation}
The e-folds number during inflation, which is defined as
\begin{equation}
\label{eq12}N=\int_{t_{hc}}^{t_{end}} H\,dt  \,,
\end{equation}
is another important parameter in inflation model. Note that the
subscript $``hc"$ shows the time of the \emph{horizon crossing} of the
physical scales and $``end"$ refers to the \emph{end of inflation}.

The important issue in studying an inflation model is its
observational viability. The Planck collaboration, in order to compare the
inflationary parameters with observational data, has expanded the scalar and tensor power spectra in a
model-independent forms as follows~\cite{pl18b,pl15}
\begin{equation}
\label{eq13} {\cal{A}}_{s}
(k)=A_{s}\left(\frac{k}{k_{*}}\right)^{n_{s}-1+\frac{1}{2}\frac{dn_{s}}{d\ln
k}\ln\big(\frac{k}{k_{*}}\big)+\frac{1}{6}\frac{d^{2}n_{s}}{d\ln
k^{2}}\ln\big(\frac{k}{k_{*}}\big)^{2}+...}\,,
\end{equation}
\begin{eqnarray}
\label{eq14} {\cal{A}}_{T}
(k)=A_{T}\left(\frac{k}{k_{*}}\right)^{n_{T}+\frac{1}{2}\frac{dn_{T}}{d\ln
k}\ln\big(\frac{k}{k_{*}}\big)+...}\,.
\end{eqnarray}
In the above equations, the amplitude of the scalar and also tensor
perturbations have been characterized by $A_{s}$ and $A_{T}$,
respectively. Also, $\frac{dn_{s}}{d\ln k}$ and $\frac{dn_{T}}{d\ln
k}$ are respectively the running of the scalar and tensor spectral
index. The term $\frac{d^{2}n_{s}}{d\ln k^{2}}$ is the running
of the running of the scalar spectral index. Another interesting and
important parameter is the tensor-to-scalar ratio which in terms of
the power spectra is given by
\begin{eqnarray}
\label{eq15}
r=\frac{{\cal{A}}_{T}(k_{*})}{{\cal{A}}_{s}(k_{*})}=16\epsilon\,.
\end{eqnarray}
In this setup one can defined the scalar and tensor spectral indices as
\begin{equation}
\label{eq16} n_{s}=1+\frac{d \ln {\cal{A}}_{s}}{d \ln
k}\Bigg|_{c_{s}k=aH}=1+(-2\epsilon-\eta-s)\,,
\end{equation}
\begin{equation}
\label{eq17} n_{T}=\frac{d \ln {\cal{A}}_{T}}{d \ln
k}\Bigg|_{c_{s}k=aH}=-2\epsilon\,,
\end{equation}
respectively. These are important parameters in studying the
inflation models and checking their observational viability. To find
an efficient way to use these parameters in inflation analysis, one can express these quantities
in terms of the e-folds number via the Hubble parameter and its
derivatives as follows.

\section{Reconstruction of the Model with e-folds Number}

In this section, we reconstruct the main equations of our model in
terms of the Hubble parameter and its derivatives, following
Refs.~\cite{Bam14,Odi15}. From now on, we assume
$f^{-1}(\phi)=V(\phi)$. By introducing a new scalar field $\varphi$,
identified by the e-folds number $N$ and characterizing the DBI
field as $\phi=\phi(\varphi)$, we can find the potential $V(\varphi)$
in terms of the Hubble parameter and its derivatives as
\begin{eqnarray}
\label{eq18} V=\frac{3}{2}\,{\frac { \left( 2\,H {\it H'
}\,{\kappa}^{2}+3\, H   ^{2} \right)  H ^{2}}{{\kappa}^{2} \left( H
{\it H' }\,{\kappa}^{2}+3\,  H  ^{2} \right) }} \,,
\end{eqnarray}
where, $H=H(N)$. By this potential, we can obtain the slow-roll
parameters as well. In this regard, the slow-roll parameters are
given by
\begin{equation}
\label{eq19}\epsilon={\frac {3\, \left( 4\,{{\it
H'}}^{3}{\kappa}^{4}+3\,  H  ^{2}{\it H'' }\,{\kappa}^{2}+15\,H {
{\it H' }}^{2}{\kappa}^{2}+18\,  H  ^{2} {\it H' } \right)
^{2}}{-32\, \left( {\it H'}\,{\kappa}^{2}+3\,H \right) ^{2} \left(
{\it H' }\,{\kappa}^{2}+\frac{3}{2}\,H \right) ^{3}{\it
H'}\,{\kappa}^{2}}} ,
\end{equation}

\begin{eqnarray}
\label{eq20}\eta=\Bigg[\big(12 \kappa^{6} H ^{2} {H'}^{3}+54
\kappa^{4} H  ^{3} {H'}^{2}+54 \kappa^{2} H^{4} {H'} \big) {H'''}\nonumber\\
-\big(36 \kappa^{6} H^{2} {H'}^{2}+99 \kappa^{4} H ^{3} {H'} +27
\kappa^{2} H^{4}\big) {H''}^{2}\nonumber\\ +\big(48 \kappa^{6} H
{H'}^{4}+126 \kappa^{4} H ^{2} {H'}^{3}+108 \kappa^{2} H ^{3}
{H'}^{2}\nonumber\\+162 H  ^{4} {H'} \big) {H''} -24 {H'}^{6}
\kappa^{6}-81 {H'}^{5} H  \kappa^{4}\nonumber\\-135 {H'}^{4} H ^{2}
\kappa^{2}-162 {H'}^{3} H  ^{3}\Bigg] \Bigg[\big(4 {H'}^{3}
\kappa^{4}+3 H ^{2} {H''} \kappa^{2}\nonumber\\+15 H {H'}^{2}
\kappa^{2}+18 H
^{2} {H'} \big) \left({H'} \,\kappa^{2}+3 H \right)\nonumber\\
\left(2 {H'} \,\kappa^{2}+3 H \right) {H'}\Bigg]^{-1} ,
\end{eqnarray}

and

\begin{equation}
\label{eq21}s=\frac{2 \kappa^{2} \left({H''} H -{H'}^{2}\right)}{H
\left(2 {H'} \,\kappa^{2}+3 H \right)}\,.
\end{equation}

By having the slow-roll parameters in terms of the Hubble parameter
and its derivatives, it is possible to study the model in the
intermediate inflation case. Before we do that, we formulate the effects
of the presence of the minimal length in a general inflation model.

\section{The Effect of the Minimal Length on the Scalar Spectral Index}

It is believed that quantum gravity implies a minimal measurable
length that modifies the uncertainty principle as~\cite{Cav03}
\begin{equation}
\label{eq22-1} \Delta X_{i}\gtrsim \frac{\hbar}{\Delta
P_{i}}\left[1+\frac{\alpha\, l_{Pl}^{2}}{\hbar^2}\left(\Delta
P_{i}\right)^{2}\right]\,,
\end{equation}
where $\alpha$ is the deformation parameter of quantum mechanics and
corresponds to the ultraviolet effect of the theory. In fact, by
considering this quantum gravitational effect, the canonical
commutation relation is modified as follows
\begin{equation}
\label{eq22}
[X_{i},P_{i}]=i\hbar(\delta_{ij}+\alpha_{ijkl}\,p^k\,p^l+...)\,.
\end{equation}
In the presence of a minimal measurable length, we define position
and momentum operators in phase space as follows
\begin{eqnarray}\label{eq23}
X^i=x^{i}\,,\quad P^i=p^i(1+\alpha p^2)\,,
\end{eqnarray}
where $x$ and $p$ are Heisenberg's algebraic position and momentum
operators in standard quantum mechanics. Given the relationship
between wave number and momentum ($p = \hbar k$), we have the
corrected wave number as follows~\cite{Taw13}
\begin{eqnarray}\label{eq24}
K^i=k^{i}(1+\alpha k^2)\,,
\end{eqnarray}
where here and in what follows we have set $\hbar=1$. Now,
due to the correction of the wave number, the definition of
the scalar spectral index is modified as follows
\begin{eqnarray}
\label{eq25} n_{s}=1+\frac{d \ln {\cal{A}}_{s}}{d \ln k(1+\alpha
k^2)}\Bigg|_{c_{s}k=aH} \nonumber\\ \cong 1+(1+\alpha k^2)\frac{d
\ln {\cal{A}}_{s}}{d \ln k}\Bigg|_{c_{s}k=aH}\nonumber\\=1+(1+\alpha
k^2)(-2\epsilon-\eta-s)\,.\hspace{0.5cm}
\end{eqnarray}
Also, the definition of the tensor spectral index is modified as
\begin{eqnarray}
\label{eq26} n_{T}=\frac{d \ln {\cal{A}}_{T}}{d \ln k(1+\alpha
k^2)}\Bigg|_{c_{s}k=aH} \nonumber\\ \cong (1+\alpha k^2)\frac{d \ln
{\cal{A}}_{T}}{d \ln k}\Bigg|_{c_{s}k=aH}=-(1+\alpha
k^2)(2\epsilon)\,.
\end{eqnarray}
So, if we believe in the presence of the natural cutoffs in early
universe, these cutoffs show themselves in the primordial
perturbation parameters explicitly. This can affect on the
formulation and numerical results of the inflation models. In this
regard, we perform numerical analysis on the DBI model in the
presence of the quantum gravitational effect encoded as a minimal
measurable length of the order of the Planck length. For this
purpose, in the what follows, we consider an intermediate scale
factor as $a=a_{0}\exp\left(b\,t^\beta\right)$ and find the
potential and the DBI tension in terms of the intermediate parameter
$\beta$. In this way, we get the scalar spectral index and
tensor-to-scalar ratio in terms of the model's parameters and
analyze the model parameter space numerically.

\section{Intermediate DBI Inflation in the Presence of a Minimal Measurable
Length and Its Observational Viability}

One of the interesting scenarios in inflation is the Intermediate
Inflation. In the intermediate inflation, the scale factor evolves
with time as
\begin{eqnarray}
\label{eq27}a=a_{0}\,\exp(b\,t^{\beta})\,,
\end{eqnarray}
where the constant parameter $\beta$ is in the range $0<\beta<1$.
Note that in intermediate inflation, the universe expands faster
than power-law inflation and slower than the exponential inflation. By
this intermediate scale factor, we get the following expression for
the Hubble parameter
\begin{eqnarray}
\label{eq28}H(N)=N\left(\frac{N}{b}\right)^{-\frac{1}{\beta}}\,\beta\,.
\end{eqnarray}
By this definition of the Hubble parameter, we can find the
slow-roll parameters in terms of the intermediate parameter $\beta$
as follows
\begin{equation}
\label{eq29}\epsilon={\frac {243 ( 1-\beta ) \left(
\frac{2{\kappa}^{4}\left( \beta-1 \right) ^{2}}{9} +\frac{5N\beta{
\kappa}^{2}}{6} \left( \beta-\frac{6}{5} \right) +{N}^{2}{\beta}^{2}
\right) ^{2}}{{\kappa}^{2} \left( 2{
\kappa}^{2}\beta+3\,\beta\,N-2{\kappa}^{2} \right) ^{3} \left( {
\kappa}^{2}\beta+3\beta\,N-{\kappa}^{2} \right) ^{2}}} ,
\end{equation}

\begin{eqnarray}
\label{eq30}\eta=\Bigg[ \left(
-24{\kappa}^{6}-81N{\kappa}^{4}-135{N}^{2}{ \kappa}^{2}-162\,{N}^{3}
\right) {\beta}^{4}\nonumber\\+ \left( 60\,{\kappa}^{6}+
171\,N{\kappa}^{4}+216{N}^{2}{\kappa}^{2} \right) {\beta}^{3}+12\,
\beta\,{\kappa}^{6}\nonumber\\-
 \left( 48{\kappa}^{6}+90N{\kappa}^{4} \right) {\beta}^{2}\Bigg]\Bigg[ \Big(
{\kappa}^{2}\beta+3\,\beta\,N-{\kappa}^{2 } \Big) \nonumber\\ \Big(
2{\kappa}^{2}\beta+3\beta N-2{\kappa}^{2}
 \Big) \Big( 4{\beta}^{2}{\kappa}^{4}+15N{\beta}^{2}{\kappa}^{
2}\nonumber\\-8\beta{\kappa}^{4}+18{N}^{2}{\beta}^{2}-18\beta{\kappa}^{
2}N+4{\kappa}^{4} \Big) \Bigg]^{-1} \,,\hspace{0.4cm}
\end{eqnarray}

\begin{eqnarray}
\label{eq31}s=2\,{\frac {{\kappa}^{2} \left( 1-\beta \right) }{
\left( 2\,N{\kappa} ^{2}+3\,{N}^{2} \right) \beta-2\,N{\kappa}^{2}}}
\,.
\end{eqnarray}

In this way, we get the scalar spectral index and tensor-to-scalar
ratio (equations (\ref{eq15}) and (\ref{eq25})) in terms of the
intermediate parameter $\beta$ and deformation parameter $\alpha$.
Now, we can perform numerical analysis on our model and by comparing
the results with observational data, we obtain some constraints on the
model's parameter space. As we mentioned in the introduction, from
Planck2018 TT, TE, EE+lowE+lensing +BAO +BK14 data, based on
$\Lambda$CDM$+r+\frac{dn_{s}}{d\ln k}$ model, the constraint on the
scalar spectral index and tensor-to-scalar ratio are as
$n_{s}=0.9658\pm 0.0038$ and $r<0.072$. Also, Planck2018 TT, TE, EE
+lowE+lensing+BK14+BAO+LIGO and Virgo2016 data gives the constraint
on the tensor spectral index as $-0.62<n_{T}<0.53$. By considering
these observational constraints, and using equations (\ref{eq15}), (\ref{eq25}),
and (\ref{eq26}), we obtain the ranges of the parameters $\alpha$
and $\beta$ leading to the observationally viable values of the
scalar spectral index, tensor-to-scalar ratio, and the tensor
spectral index. The results are shown in figure 1. Another way to
analyze the observational viability of the model is the study of
$r-n_s$ and $r-n_T$ behaviors. To this end, we plot $r-n_{s}$
behavior in the background of the Planck2018 TT, TE, EE +lowE
+lensing +BAO+ BK14 data in figure 2. We also plot $r-n_{T}$
behavior in the background of the Planck2018 TT, TE, EE
+lowE+lensing+BK14 +BAO+LIGO $\&$ Virgo2016 data in figure 3.
Figures 2 and 3 have been plotted for $0.6<\beta<1$ and $\alpha<
10^{50}$. Our numerical analysis shows that the intermediate DBI
model in the presence of a minimal length is consistent with the
both Planck2018 TT, TE, EE +lowE +lensing +BAO+ BK14 data and
Planck2018 TT, TE,EE +lowE+lensing+BK14 +BAO+LIGO $\&$ Virgo2016
data if $0.93 \leq \beta <1 $.

\begin{figure}
\begin{center}\includegraphics{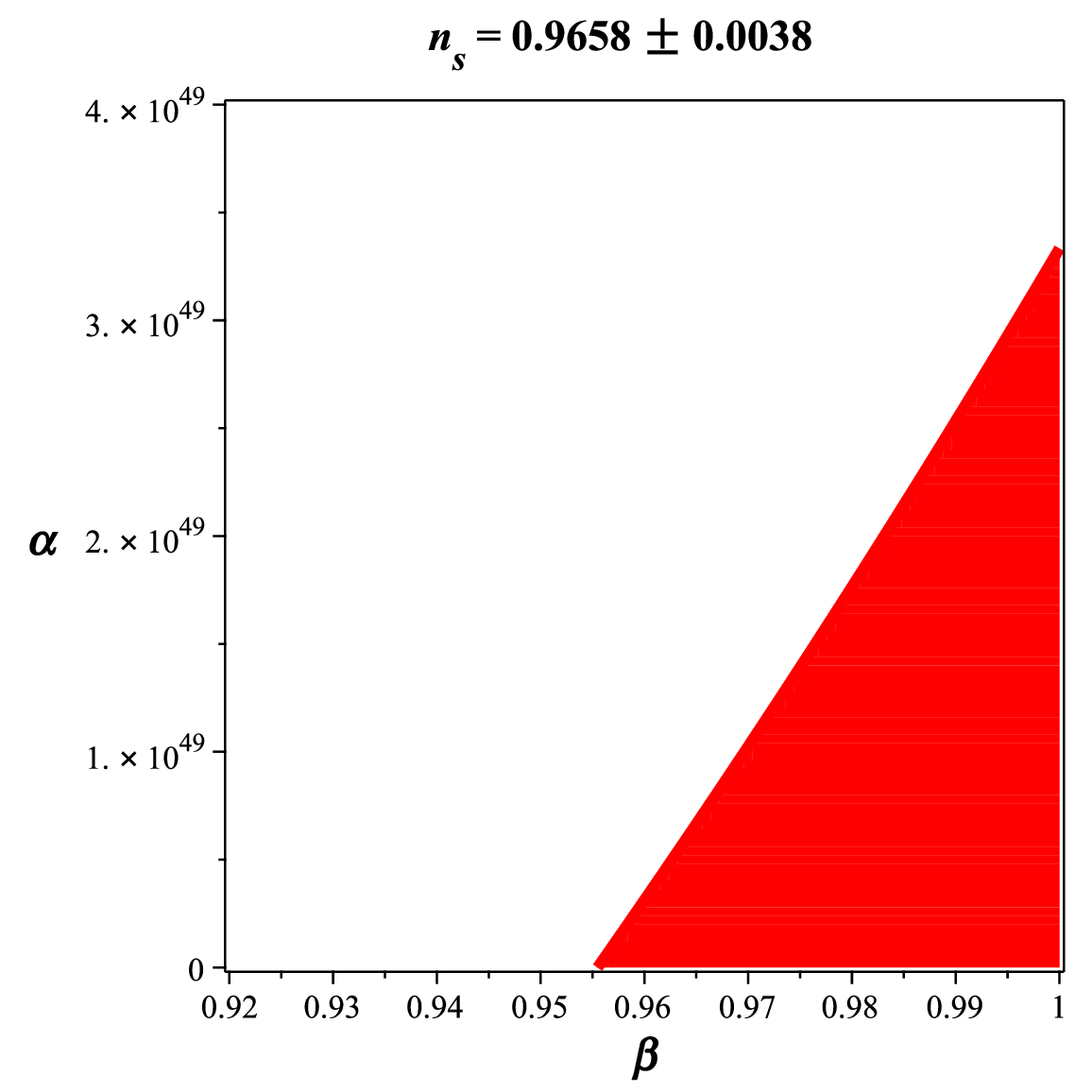}\includegraphics{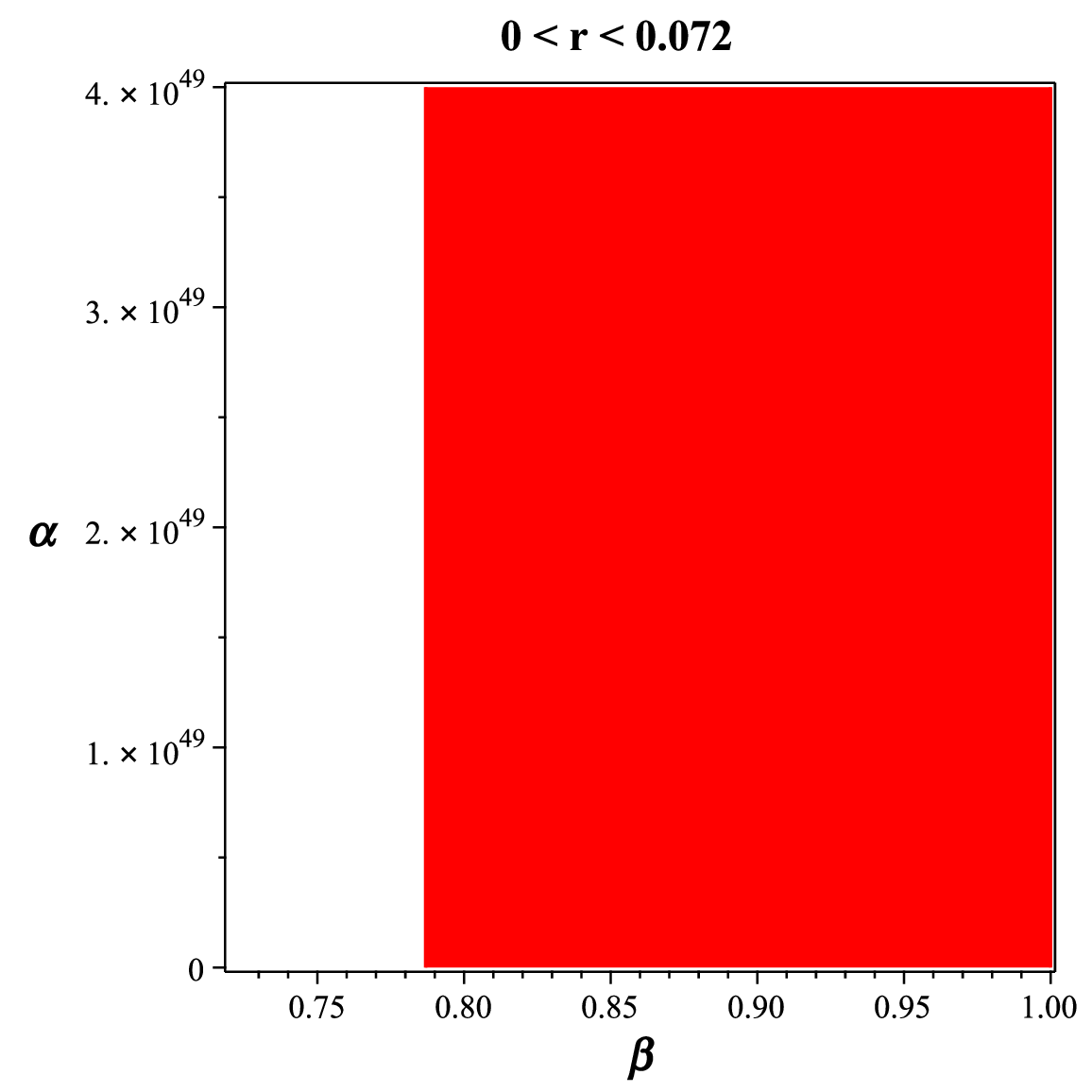}\includegraphics{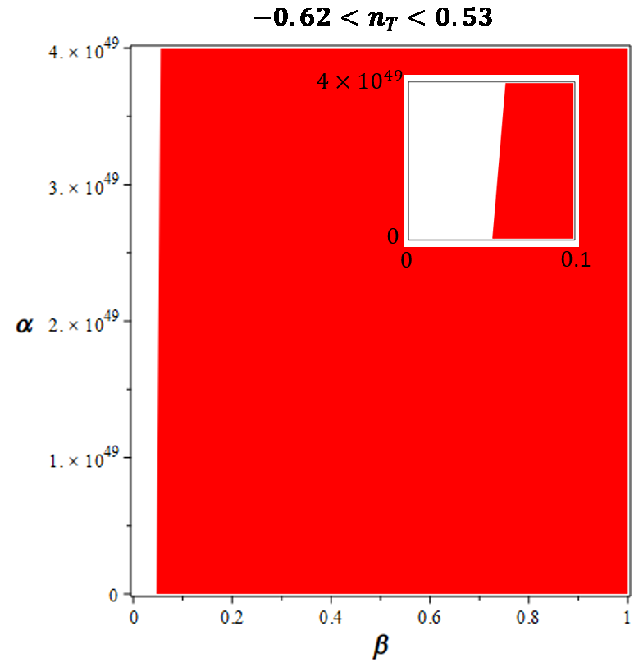} \vspace{18.5cm}
\end{center}
\caption{\label{fig3}\small {Ranges of the parameters $\alpha$ and
$\beta$ leading to the observationally viable values of the scalar
spectral index (upper-left panel), tensor-to-scalar ratio
(upper-right panel), and the tensor spectral index (lower panel) in
the intermediate DBI model in the presence of a minimal measurable
length. }}
\end{figure}

\begin{figure}
\begin{center}\includegraphics{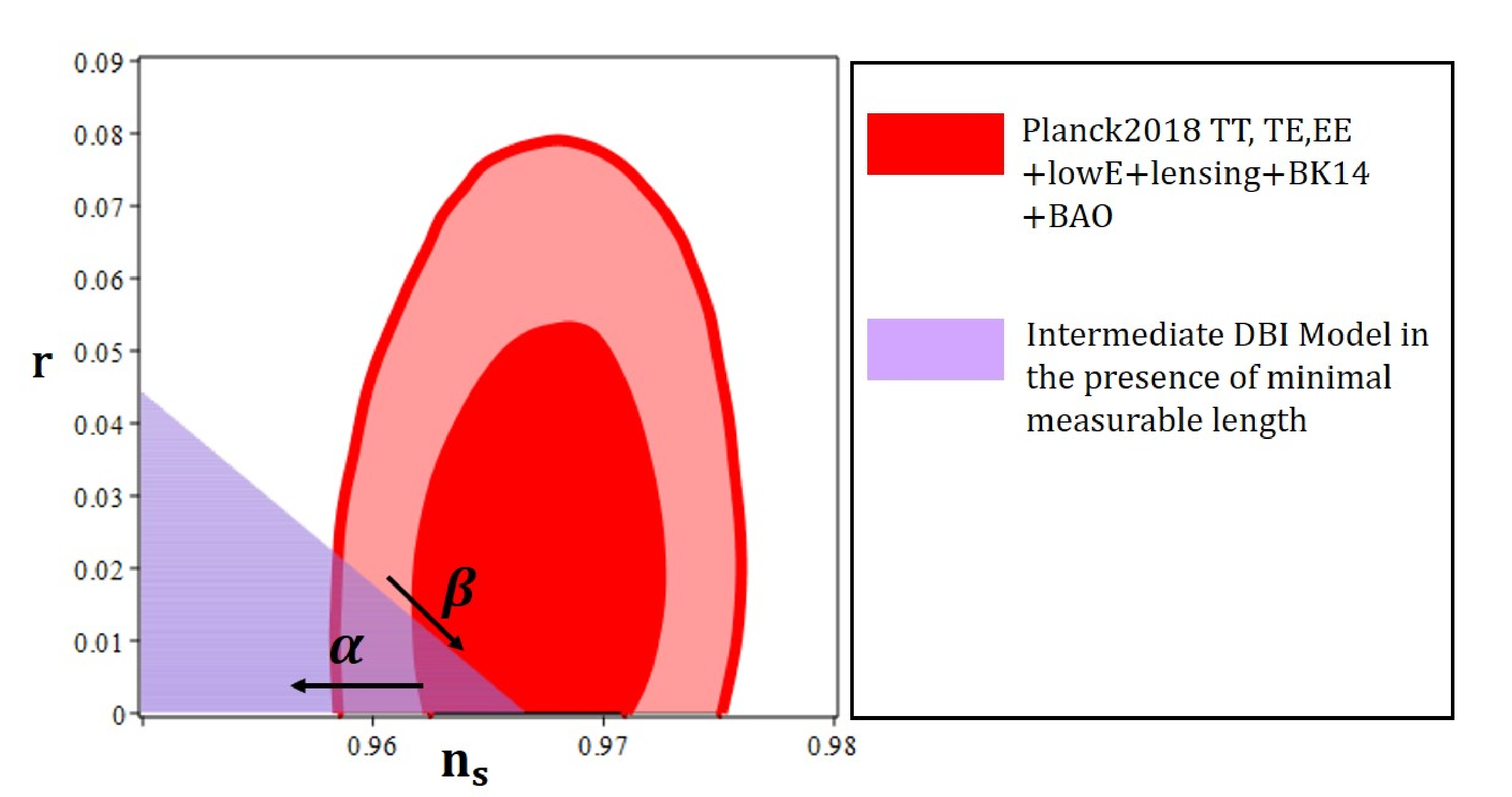} \vspace{6.5cm}
\end{center}
\caption{\label{fig3}\small {Tensor-to-scalar ratio versus the
scalar spectral index for the intermediate DBI model in the presence
of a minimal measurable length, in the background of Planck2018 TT,
TE, EE +lowE+lensing+BK14 +BAO data. The arrows show the direction
where the parameters increase.}}
\end{figure}

\begin{figure}
\begin{center}\includegraphics{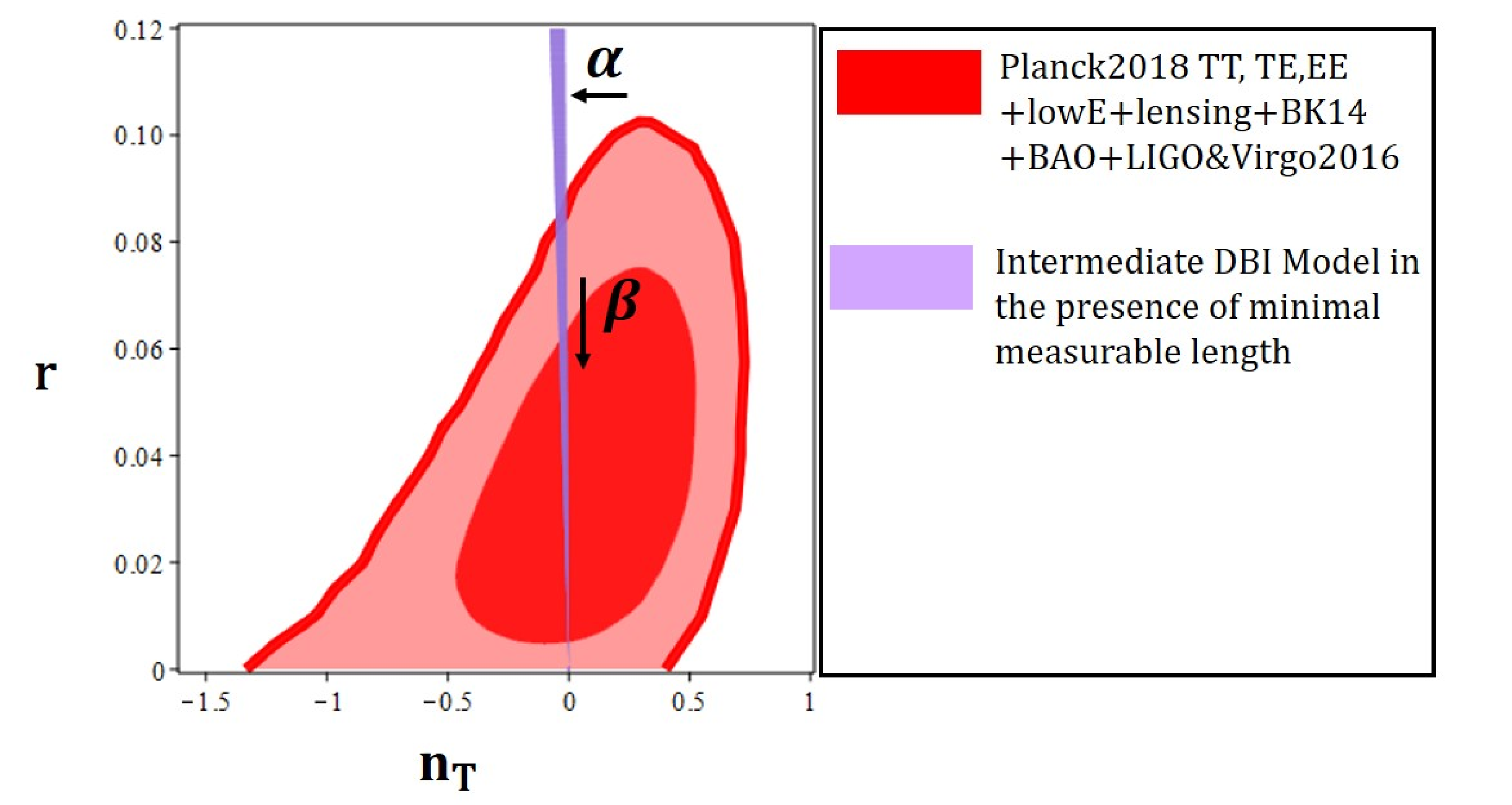} \vspace{6.5cm}
\end{center}
\caption{\label{fig3}\small {Tensor-to-scalar ratio versus the
tensor spectral index for the intermediate DBI model in the presence
of a minimal measurable length, in the background of Planck2018 TT,
TE, EE +lowE+lensing+BK14 +BAO+LIGO $\&$ Virgo2016 data. The arrows
show the direction where the parameters increase.}}
\end{figure}

To find some specified constraints on the model's parameters, we
have adopted some sample values of the intermediate parameter as
$\beta=0.94,\, 0.96$, and $0.98$. By these adopted values of
$\beta$, we have obtained some constraints on the parameter
$\alpha$, in comparison with the Planck2018 TT, TE, EE +lowE
+lensing +BAO+ BK14 data and Planck2018 TT, TE,EE +lowE+lensing+BK14
+BAO+LIGO $\&$ Virgo2016 data at $68\%$ CL and $95\%$ CL. The
results are summarized in Table 1. Note that, although by analyzing
the tensor spectral index we find the upper bound on the deformation
parameter $\alpha$ to be of the order of $10^{53}$, for some domains
within this range the scalar spectral index is not observationally
viable. In this regard, to get the observationally viable range of
$\alpha$, we should consider all three perturbation parameters
(scalar spectral index, tensor spectral index, and tensor-to-scalar
ratio). In this way, from our analysis, we find that the
observationally viable upper bound on the deformation parameter is
of the order of $10^{48}$ at $68\%$ CL and $10^{49}$ at $95\%$ CL.
This is consistent with other studies, see for
instance~\cite{Roushan19}.

\begin{table*}
\tiny \caption{\small{\label{tab:1} The ranges of the parameter
$\alpha$ in which the tensor-to-scalar ratio and the scalar spectral
index of the intermediate DBI model in the presence of a minimal
measurable length are consistent with different data sets.}}
\begin{center}
\begin{tabular}{cccccc}
\\ \hline \hline \\ & Planck2018 TT,TE,EE+lowE & Planck2018 TT,TE,EE+lowE&Planck2018 TT,TE,EE+lowE&Planck2018 TT,TE,EE+lowE
\\
& +lensing+BK14+BAO &
+lensing+BK14+BAO&lensing+BK14+BAO&lensing+BK14+BAO
\\
&  & &+LIGO$\&$Virgo2016 &LIGO$\&$Virgo2016
\\
\hline \\$\beta$& $68\%$ CL & $95\%$ CL &$68\%$ CL & $95\%$ CL
\\
\hline\hline \\  $0.94$ & not consistent & $0<
\alpha<1.15\times10^{49}$ & $0<\alpha<5.18\times10^{52}$ & $0<\alpha<1.05\times10^{53}$\\ \\
\hline
\\$0.96$& $0<
\alpha<5.04\times10^{48}$ & $0< \alpha<2.67\times10^{49}$
&$0<\alpha<7.15\times10^{52}$ & $0<\alpha<1.80\times10^{53}$
\\ \\ \hline\\
$0.98$& $0< \alpha<1.84\times10^{48}$ & $0<
\alpha<4.28\times10^{49}$ & $0<\alpha<7.85\times10^{52}$ & $0<\alpha<4.29\times10^{53}$ \\ \\
\hline \hline
\end{tabular}
\end{center}
\end{table*}

\section{Summary and Conclusion}

In this work, we have studied the intermediate DBI inflation in the
presence of a natural cutoff as a minimal measurable length. By
reviewing the DBI model, we have presented the main equations of the
model and also some important inflation observables such as the scalar
spectral index, tensor spectral index and the
tensor-to-scalar ratio. After that, we reconstructed the model with
e-folds number and re-formulated the slow-roll parameters in terms
of the Hubble parameter and its derivatives. In this way, we have
prepared ourselves to consider the intermediate scale factor in the
DBI inflation model. Then, by considering that the presence of a
minimal measurable length modifies the momentum operator, we have
presented the modified form of the wave number. This is an important
point in our setup. The reason is that to obtain the scalar and tensor spectral
indices, the wave number plays a crucial role. In this regard, we
have obtained these parameters by using the modified wave number and
in this way, the effects of a minimal measurable length have been
included in the perturbation parameters $n_s$ and $n_{T}$. Then, we
considered an intermediate scale factor and obtained the slow-roll
parameters in terms of the intermediate parameter. In this way, the
perturbation parameters have been expressed in terms of the deformation and
intermediate parameters. Finally, we have performed a numerical
analysis on the model's parameters and compared the results with
Planck 2018 different joint data sets. In this regard, by using the
observational constraints on the perturbation parameters, implied by
Planck2018 TT, TE, EE +lowE +lensing +BAO+ BK14 data and Planck2018
TT, TE,EE +lowE+lensing+BK14 +BAO+LIGO $\&$ Virgo2016 data, we have
obtained the ranges of the deformation parameter $\alpha$ and
intermediate parameter $\beta$, leading to observationally viable
values of $r$, $n_{s}$, and $n_{T}$. We have also explored $r-n_{s}$
behavior in the background of Planck2018 TT, TE, EE +lowE +lensing
+BAO+ BK14 data and $r-n_{T}$ behavior in the background of
Planck2018 TT, TE,EE +lowE+lensing+BK14 +BAO+LIGO $\&$ Virgo2016
data at $68\%$ CL and $95\%$ CL. By adopting some sample viable values of
$\beta$ as $\beta=0.94,\, 0.96$, and $0.98$, we have obtained some
constraints on the deformation parameter $\alpha$. According to our
numerical analysis, the intermediate DBI inflation model in the
presence of a minimal measurable length is observationally viable if
the upper bound on the deformation parameter is of the order of
$10^{48}$ at $68\%$ CL and $10^{49}$ at $95\%$ CL.\\

{\bf Acknowledgement}\\

We thank the referee for the insightful comments that have
improved the quality of the paper.\\
This research work has been supported by a research grant from the University of Mazandaran.\\

\textbf{Data availability statement} All data generated or analyzed
during this study are included in this article.


\begin{thebibliography}{0}

\bibitem{Gut81} \Name{A. Guth} \REVIEW{Phys. Rev. D} {23}{1981}{347}.

\bibitem{Lin82} \Name{A. D. Linde} \REVIEW{Phys. Lett. B} {108}{1982}{389}.

\bibitem{Alb82} \Name{A. Albrecht \and P. Steinhard} \REVIEW{Phys. Rev. D} {48}{1982}{1220}.

\bibitem{Lin90} \Name{A. D. Linde} \Book{Particle Physics and Inflationary Cosmology}
\Publ{Harwood Academic Publishers, Chur, Switzerland} \Year{1990}.

\bibitem{Lid00a} \Name{A. Liddle \and D. Lyth} \Book{Cosmological Inflation and Large-Scale Structure}
\Publ{Cambridge University Press} \Year{2000}.

\bibitem{Lid97} \Name{J. E. Lidsey, A. R. Liddle, E. W. Kolb, E. J. Copeland, T. Barreiro \and M. Abney} \REVIEW{Rev. Mod. Phys.}
{69} {1997}{373}.

\bibitem{Rio02} \Name{A. Riotto} [arXiv:hep-ph/0210162] (2002).

\bibitem{Lyt09} \Name{D. H. Lyth \and A. R. Liddle} \Book{The Primordial Density Perturbation} \Publ{Cambridge University Press} \Year{2009}.

\bibitem{Mal03} \Name{J. M. Maldacena} \REVIEW{JHEP} {0305}{2003}{013}.

\bibitem{Sil04} \Name{E. Silverstein \and  D. Tong} \REVIEW{Phys. Rev. D} {70}{2004}{103505}.

\bibitem{Che07} \Name{X. Chen, M.-x. Huang, S. Kachru \and G. Shiu} \REVIEW{JCAP} {0701}{2007}{002}.

\bibitem{Ali04} \Name{M. Alishahiha, E. Silverstein \and D. Tong} \REVIEW{Phys. Rev. D} {70}{2004}{123505}.

\bibitem{Che05} \Name{X. Chen} \REVIEW{J. High Energy Phys.} {0508}{2005}{045}.

\bibitem{Noz13} \Name{K. Nozari \and N. Rashidi} \REVIEW{Phys. Rev. D} {88}{2013}{084040}.

\bibitem{Amani18} \Name{R. Amani, K. Rezazadeh, A. Abdolmaleki \and Kayoomars Karami} \REVIEW{AstroPhys. J.} {853}{2018}{188}.

\bibitem{Rasouli19} \Name{S. Rasouli, K. Rezazadeh, A. Abdolmaleki \and K. Karami} \REVIEW{Eur. Phys. J. C} {79}{2019}{79}.

\bibitem{Noz19} \Name{K. Nozari \and N. Rashidi} \REVIEW{The Astrophysical Journal} {882}{2019}{78}.

\bibitem{Ras21a} \Name{N. Rashidi \and K. Nozari} \REVIEW{Eur. Phys. J. C} {81}{2021}{834}.

\bibitem{Duf97} \Name{M. J. Duff} \REVIEW{Nucl. Phys. B} {125}{1997}{334}.

\bibitem{Noj02} \Name{S. Nojiri \and S. D. Odintsov} \REVIEW{IJMPA} {17}{2002}{4809}.

\bibitem{Noj03} \Name{S. Nojiri \and S. D. Odintsov} \REVIEW{Phys. Lett. B} {571}{2003}{1}.

\bibitem{Ras21} \Name{N. Rashidi} \REVIEW{The Astrophysical Journal} {914}{2021}{29}.

\bibitem{Venez86} \Name{G. Veneziano} \REVIEW{EPL} {2}{1986}{3}.

\bibitem{Amati87} \Name{D. Amati, M. Cialfaloni, \and G. Veneziano} \REVIEW{Phys. Lett. B} {197}{1987}{81}.

\bibitem{Gross87} \Name{D. J. Gross \and P. F. Mende} \REVIEW{Phys. Lett. B} {197}{1987}{129}.

\bibitem{Amati89} \Name{D. Amati, M. Cialfaloni, \and G. Veneziano} \REVIEW{Phys. Lett. B} {216}{1989}{41}.

\bibitem{Konishi90} \Name{K. Konishi, G. Paffuti, \and P. Provero} \REVIEW{Phys. Lett. B} {234}{1990}{276}.

\bibitem{Mag93} \Name{M. Maggiore} \REVIEW{Phys. Lett. B} {304}{1993}{65}.

\bibitem{Capo2000} \Name{S. Capozziello, G. Lambiase, \and G. Scarpetta} \REVIEW{IJTP} {39}{2000}{15}.

\bibitem{Magg93} \Name{M.Maggiore} \REVIEW{Phys. Lett. B} {319}{1993}{83}.

\bibitem{Mag94} \Name{M. Maggiore} \REVIEW{Phys. Rev. D} {49}{1994}{5182}.

\bibitem{Hossen2012} \Name{S. Hossenfelder} \REVIEW{CQG} {29}, {2012}{115011}.

\bibitem{Hossen2013} \Name{S. Hossenfelder} \REVIEW{Living Rev. Relativ} {16}{2013}.

\bibitem{Scar1999} \Name{F. Scardigli} \REVIEW{Phys. Lett. B} {452}{1999}{39}.

\bibitem{Kempf95} \Name{A. Kempf, G. Mangano \and R. B. Mann} \REVIEW{Phys. Rev. D} {52}{1995}{1108}.

\bibitem{Nozari12b} \Name{K. Nozari \and A. Etemadi} \REVIEW{Phys. Rev. D}  {85}{2012}{104029}.

\bibitem{Roushan14} \Name{M. Roushan \and K. Nozari} \REVIEW{AHEP}{2014}{Article ID 353192}.

\bibitem{Roushan16} \Name{K. Nozari \and M. Roushan} \REVIEW{IJGMMP} {13}{2016}{1650054}.

\bibitem{Roushan18} \Name{M. Roushan \and K. Nozari} \REVIEW{IJGMMP} {15}{2018}{1850136}.

\bibitem{Roushan19} \Name{M. Roushan \and K. Nozari} \REVIEW{EPJC} {79}{2019}{212}.

\bibitem{pl18a} \Name{N. Aghanim, Y. Akrami, M. Ashdown, et al.} [arXiv:1807.06209] (2018).

\bibitem{pl18b} \Name{Y. Akrami, F. Arroja, M. Ashdown, et al.} [arXiv:1807.06211] (2018).

\bibitem{pl15} \Name{P. A. R. Ade, M. Arnaud, M. Ashdown, J. Aumont, C. Baccigalupi,
\emph{et al.}} \REVIEW{Astron. Astrophys.} {594}{2016}{A20}.

\bibitem{Bam14} \Name{K. Bamba, S. Nojiri, S. D. Odintsov \and D. S\'{a}ez-G\'{o}mez} \REVIEW{Phys. Rev. D} {90}{2014}{124061}.

\bibitem{Odi15} \Name{S. D. Odintsov \and  V. K. Oikonomou} \REVIEW{Annals of Physics} {363}{2015}{503}.

\bibitem{Cav03} \Name{M. Cavaglia, S. Das \and R. Maartens} \REVIEW{Class. Quant. Grav} {20}{2003}{L205}.

\bibitem{Taw13} \Name{A. Tawfik, H. Magdy \and A. Farag Ali} \REVIEW{Gen. Rel. Grav} {45}{2013}{1227}.



\end{thebibliography}
\end{document}